\documentclass[runningheads]{llncs}
\usepackage{float}
\usepackage[hidelinks]{hyperref}
\usepackage[T1]{fontenc}
\usepackage{mathrsfs}
\usepackage{threeparttable}
\usepackage{diagbox}
\usepackage{multirow}
\usepackage{booktabs}
\usepackage{algorithmic}
\usepackage[ruled,noend,linesnumbered]{algorithm2e}
\usepackage{url}
\usepackage{hyperref}
\usepackage{bm}
\usepackage{todonotes}
\usepackage{amsmath}
\usepackage{enumitem}
\usepackage{graphicx}
\usepackage{multirow}
\usepackage{array}
\usepackage{caption}
\usepackage[utf8]{inputenc}
\usepackage{subcaption}
\usepackage{xcolor}
\usepackage{tikz}
\usetikzlibrary{matrix}
\usepackage{pgfplots}
\usepackage{pgfplotstable}
\usetikzlibrary{patterns}
\usepackage{centernot}
\pgfplotsset{compat=newest} 
\usepackage[hyphenbreaks]{breakurl}
\usepackage{graphicx}
\usepackage{amsfonts}
\usepackage{algorithm2e}
\begin{document}

\title{Sync Without Guesswork: Incomplete Multivariate Time Series Alignment}

%
%
%
\author{
Ding Jia\inst{1} \and
Jingyu Zhu\inst{1} \and
Yu Sun\inst{1} \and
Aoqian Zhang\inst{2} \and
Shaoxu Song\inst{3} \and
Haiwei Zhang\inst{1} \and
Xiaojie Yuan\inst{1}
}

\authorrunning{D. Jia et al.}

\institute{
Nankai University, Tianjin, China\\
\email{\{2310047,2120240709\}@mail.nankai.edu.cn,
sunyu@nankai.edu.cn,
\{zhhaiwei,yuanxj\}@nankai.edu.cn}
\and
Beijing Institute of Technology, Beijing, China\\
\email{aoqian.zhang@bit.edu.cn}
\and
Tsinghua University, Beijing, China\\
\email{sxsong@tsinghua.edu.cn}
}
\maketitle              
\begin{abstract}
Multivariate time series alignment is critical for ensuring coherent analysis across variables, but missing values and timestamp inconsistencies make this task highly challenging. 

Existing approaches often rely on prior imputation, which can introduce errors and lead to suboptimal alignments. To address these limitations, we propose a constraint-based alignment framework for incomplete multivariate time series that avoids imputation and ensures temporal and structural consistency. We further design efficient approximation algorithms to balance accuracy and scalability. Experiments on multiple real-world datasets demonstrate that our approach achieves superior alignment quality compared to existing methods under varying missing rates.

Our contributions include: (1) formally defining incomplete multiple temporal data alignment problem; (2) proposing three approximation algorithms balancing accuracy and efficiency; and (3) validating our approach on diverse real-world datasets, where it consistently outperforms existing methods in alignment accuracy and the number of aligned tuples.

\end{abstract}

\section{INTRODUCTION}
Multivariate time series is fundamental to applications from sensor monitoring to financial forecasting. However, real-world multivariate time series are often collected separately (e.g., fuel consumption \cite{intro_4} and voltage prediction \cite{intro_5}), which brings two key quality issues: timestamp inconsistency and missing data.

Owing to transmission or network latency, values corresponding to different variables that are generated simultaneously may be recorded with misaligned timestamps \cite{intro_1}. This phenomenon, often referred to as timestamp inconsistency, is commonly observed in time-sensitive systems \cite{SAMC,intro_2}.

In addition, collecting multivariate time series data across different domains faces challenges. Therefore, datasets typically exhibit substantial missing data, which can severely compromise the accuracy and reliability of subsequent analyses. For instance, the widely used medical time series dataset PhysioNet2012 reports a high average missing rate, making robust analysis difficult \cite{intro_3,mobile,Imputation}.

Most existing alignment methods, including variants of dynamic time warping (e.g., DTW \cite{DTW}, CTW \cite{CTW}) and similarity-based approaches (e.g., SAMC \cite{SAMC}), assume complete sequences. When applied to incomplete data, existing methods resort to imputation techniques such as linear interpolation or model-based filling. However, performing imputation and alignment as separate steps introduces two key limitations. First, errors from imputation propagate to the alignment stage, compounding inaccuracies. Second, imputation models—even when leveraging multiple dependencies—typically estimate each missing value in isolation without considering the structural constraints imposed by alignment. As a result, the imputed values may be locally plausible but globally inconsistent, leading to aligned tuples that violate temporal or statistical dependencies inherent in the original data. This motivates a joint formulation that enforces alignment constraints during tuple selection rather than relying on independent imputation.

To address these challenges, we propose a constraint-based alignment method for incomplete multivariate time series without prior imputation, while ensuring that the aligned tuples are not only temporally coherent but also consistent with the underlying time series dynamics.

Our contributions are as follows:

(1) We define the alignment task with three main constraints to keep the results consistent and reasonable: the time constraint to make sure aligned values are close in time, the position constraint to handle missing timestamps by limiting index gaps, and the model consistency constraint to ensure the aligned tuples statistically valid. Our objective is to maximize the total weight of aligned tuples, balancing completeness and positional fidelity.
(2) We design efficient algorithms with tunable trade-offs: a combinatorial method with exponential complexity, an approximation based on weighted k-set packing \cite{BestImp} with polynomial complexity, and two practical algorithms based on greedy and expectation strategies with near-linear complexity, offering flexibility between alignment quality and computational efficiency.

(3) We provide theoretical guarantees on the approximation ratios of all proposed algorithms.
(4) Experiments on four real-world datasets with 10\%\textasciitilde 40\% missing rates show that our method consistently outperforms six existing baselines, and remains robust across varying data sizes and missing patterns. The code is in the Github repository\footnote{https://anonymous.4open.science/r/Incomplete-Multivariate-Time-Series-Alignment-8EC0}.

\section{PROBLEM STATEMENT}
In this section, we first introduce the core constraints and weight definitions that govern our alignment process. These constraints ensure that the resulting aligned tuples are not only temporally and structurally consistent but also compatible with the underlying time series model. By quantifying tuple quality through a weight function, we create a basis for selecting high-quality alignments in subsequent algorithmic steps.

\subsection{Constraints}

Traditional methods that rely on exact timestamp matching result in very limited usable data. Therefore, to address this limitation, we introduce a time constraint that permits the alignment of variable values recorded within a short time window. This relaxation increases the number of possible tuples while still preserving temporal coherence. Following SAMC \cite{SAMC}, we use timestamp difference constraint to restrict timestamps in all the aligned tuples.

Assuming we have $m$ time series to be aligned, represented by timestamps $T\in\mathbb{R}^{n\times m}$ and corresponding values $V\in\mathbb{R}^{n\times m}$, where each row contains a timestamp–value pair $(T_{i}, V_{i})$ from the $i$-th time series. The aligned result is denoted as $R_m$.

Each tuple $r$ in aligned result $R_m$ is denoted as $\left \{(T_1,V_1),(T_2,V_2)...(T_m,V_m)\right \}$. $T$ and $V$ are the timestamps and values of each tuple.
\begin{definition}
The timestamp similarity of each aligned tuple $r\in R_m$ is defined on the difference between any two non-missing timestamps from it.
\end{definition}
\begin{equation}
\forall i,j \in[1,m], \Theta(r) =max(\left| T_{i}-T_{j} \right|).
\end{equation}
Each $r\in R_m$ is expected to be aligned in a short period, so their timestamps should have distances no greater than a threshold $\theta$.
\begin{definition}\label{def:time}
We say that an aligned tuple $r$ satisfies the time constraint if $\Theta(r)\leq \theta$.
\end{definition}
To handle the missing timestamps in time series, we propose position constraint to restrict the line indicies distances in each $r \in R_m$.
\begin{definition}
The position similarity of each aligned tuple $r\in R_m$ is defined on the difference between line indicies of $(T,V)$.
\end{definition}
\begin{equation}
\forall i,j \in[1,m],\Phi (r)=max(\left| I[(T_i,V_i)]-I[(T_j,V_j)] \right|).
\end{equation}

$I[(T,V)]$ denotes the row index of the timestamp–value pair $(T_i,V_i)$ in the original data. Each $r\in R_m$ is expected to be aligned in a short distance, so their indicies should have differences no greater than a threshold $\beta$.
\begin{definition}\label{def:position}
We say that an aligned tuple $r$ satisfies the position constraint if $\Phi(r)\leq \beta$.
\end{definition}

To guarantee the consistency between each other in the aligned result, we use time series model to predict its values and evaluate its consistency with its loss.

Firstly, we have $A_{i,j} = |M_{i,j} - V_{i,j}| \times mask_{i,j},i\in [1,n], j\in [1,m]$, $A_{i,j}$ is the loss of $i$-th value in the $j$-th time series. $M_{i,j}$ is the predicted value of $V_{i,j}$ by time series model. $mask_{i,j}$ stands for the missing situation of $V_{i,j}$, which equals to 0 when it is missing otherwise 1. Then, we have $\mu_{j} = F_j\times (V_{j}^{max} -  V_{j}^{min}),j\in [1,m]$, $V_j$ is the $j$-th time series and $F_j$ is the total number of non-missing of the $j$-th time series. And we have $L_{j} = \sum_{i=1}^{|R_m|}\frac{A_{i,j}}{\mu_j},j\in [1,m]$, $L_j$ is the normalized mean absolute error of the non-missing values in the $j$-th time series.
\begin{definition}
For each aligned result $R_m$ and trained time series model $M$, its consistency is defined as:
\end{definition}
\begin{equation}
\Delta(R_m,M) = \frac{1}{m} \sum_{j=1}^{m} L_j.
\end{equation}

The smaller the $\Delta(R_m,M)$ is, the more the aligned result is consistent within.
\begin{definition}\label{def:model}
We say that an aligned result satisfies the model constraint if $\Delta(R,M)\leq \delta$.
\end{definition}

\subsection{Weight Definition}

To make the number of non-missing aligned pairs $\{(T_i,V_i),(T_j,V_j)\},i,j\in[1,m]$ in aligned result as large as possible, we value each tuple with their weights. 
\begin{definition}\label{def:weight}
For each $r\in R_m$, we define its weight $W(r)$ by the sum of pairs $p(r)$ and distances $d(r)$ in it.
\end{definition}
\begin{align}
\left\{
\begin{aligned}
&p(r)=\frac{\lambda \times (\lambda-1)}{2}, \\
&d(r)=\sum_{i=1}^{m}\sum_{j=i+1}^{m}\left |I[(T_i,V_i)]-I[(T_j,V_j)] \right |,i,j\in [1,m],\\
&W(r)=\frac{k1 \times p(r)+b}{k2 \times d(r)+c}.
\end{aligned}
\right.
\end{align}
$\lambda$ is the total number of non-missing values in $r$.

Here, $k_1$ and $k_2$ are coefficients that control the importance of value completeness and positional compactness, respectively, while $b$ and $c$ are bias terms introduced to ensure numerical stability and prevent zero-division. These parameters are predefined to flexibly balance the trade-off between the number of valid pairs and their positional coherence within each tuple, making the weight function adaptable to different data characteristics.
\subsection{Problem Statement}\label{problem}
To formally address the incomplete multivariate time series alignment problem, we define a set of constraints that each aligned tuple must satisfy. Unlike existing methods that rely heavily on imputation—often introducing noise and inconsistency—our formulation incorporates structural constraints directly into the alignment process. This enables more reliable tuple selection while ensuring temporal coherence, positional continuity, and model-level consistency.
\begin{problem}[Constraint-based Incomplete Time Series Alignment]\label{problem statement}
Each aligned tuple $r\in R_m$ should satisfies:
(1) each $(T,V)$ can only be aligned once, \emph{i.e.}, $\forall r_i,r_j\in R_m,k\in [1,m],r_i[k]\ne r_j[k]$, $r[k]$ is the $k$-th $(T,V)$ in tuple $r$.
(2) $\forall r\in R_m, \Theta(r)\leq\theta$,
(3) $\forall r\in R_m, \Phi(r)\leq\beta$,
(4) $\Delta(R_m,M)\leq \delta$,
(5) the sum of weight is maximized:
\begin{equation}
\max_{R_m} \sum_{r \in R_m} W(r).
\end{equation}
\end{problem}

$\sum_{r \in R_m} W(r)$ denotes the total weight of tuples in $R_m$. If tuple $r_i$ and $r_j$ have no $(T,V)$ aligned repeatedly, it is denoted as $r_{i} \centernot\asymp r_{j}$. Otherwise, we say $r_i$ and $r_j$ conflict with each other, denoted as $r_{i} \asymp r_{j}$.

This constraint-based formulation not only avoids the drawbacks of direct imputation but also ensures that the aligned results are both structurally and semantically consistent with the underlying time series model.
\section{ALIGNMENT ALGORITHM}\label{exact}
In our alignment process, motivated by SAMC \cite{SAMC}, we perform alignment by selecting candidate tuples that satisfy predefined constraints, composing time series and check whether each of them satisfies the model constraint.

Differently, SAMC primarily focuses on sequentially scanning and matching tuples under fixed constraint settings, while our method introduces a weight-maximization objective and integrates model-based constraints, enabling the selection of high-quality alignments even in the presence of missing or noisy data. This design allows our approach to retain the efficient conflict-resolution mechanism while addressing the limitations of purely sequential strategies.

We first generate a tuple set $R_c$ as candidate to ensure that each tuple $r \in R_c$ satisfies the time and position constraints defined in Definition~\ref{def:time} and Definition~\ref{def:position}. This candidate set serves as a feasible search space, filtering out combinations that inherently violate alignment constraints and would lead to invalid or low-quality results.

By constructing $R_c$ in advance, we reduce the search space and computational cost of subsequent optimization, while ensuring that all selected tuples are temporally and positionally coherent. Based on $R_c$, we then select a subset of tuples and compose a new aligned time series that maximizes the sum of weights, subject to the model constraint and uniqueness constraint described in Section~\ref{problem} (3) and (4).

In this section, we first introduce a pruned candidate generation method to efficiently construct $R_c$ based on time and position constraints. Then, we propose a time series composing algorithm that selects tuples from $R_c$ following the requirement described in Section~\ref{problem} (1) to form a final aligned result.
\subsection{Candidate Generation}\label{candidate}

We initialize an index set $I_c$ with $m$ numbers in it, which stands for the current line indices during the candidate searching process. For each round, the corresponding tuple will be put into candidate set $R_c$ if the two constraints are fulfilled, and move the last index in $I_c$ to the next $(T,V)$ in the same time series in order to make the tuples in candidate sorted in ascending order. If the index is out of range, we move the formal index and initialize all the index back of it with the position constraint instead of to the beginning of this column.

To prune the searching procedure, if any of the time constraints is not satisfied, considering the monotonically increasing feature of timestamps in each time series, we directly move the further back one between the $I_{c}^{max}$ or $I_{c}^{min}$ to the next line, $I_{c}^{max}$ denotes the index with the maximal timestamp in $I_c$ and $I_{c}^{min}$ denotes the index with minimal timestamp. After the moving process, we initialize the indices back of it in the same way as we mentioned above. This strategy can also be used similarly when the position constraint is not satisfied due to the same feature. Time complexity of candidate generation in the worst situation is $O(nm(\beta+1)^{m})$.

\subsection{Time Series Compose}

Based on the generated candidate set $R_c$, we iteratively select tuples to compose a new aligned time series $r_a$ following requirement (1) in Problem \ref{problem statement}, which is then added to the alternative series set $R_a$ for further evaluation. The optimization objective of this method is denoted as $W_O$.

Finally, considering the missing values in time series, we use MARSS \cite{MARSS} time series model with missing tolerance to predict each $r_a$, calculate the average NMAE and choose the one with the greatest sum of weights among those can fulfill the Definition \ref{def:model} as the final aligned result $R_m$.

$R_c[i]$ is the $i$-th candidate tuple in $R_c$. In Algorithm \ref{pseudo-exact} Line \ref{alg1:ini_s}-\ref{alg1:ini_e}, we initialize $\psi$ as a mark and $W$ as the temporal largest sum of weights. In Line \ref{alg1:traverse}, all the composed situation is traversed. In Line \ref{alg1:compose_s}-\ref{alg1:compose_e}, we compose time series, and in Line \ref{alg1:check_s}-\ref{alg1:check_e} we check if it satisfies requirement (1) in Section \ref{problem}. Finally, in Line \ref{alg1:ans_s}-\ref{alg1:ans_e}, we find the composed result with the largest sum of weights.

\begin{algorithm}[ht]
\caption{Time Series Compose($R_c$, $\delta$)}\label{pseudo-exact}
\KwIn{Candidate set $R_c$, model constraint $\delta$}
\KwOut{Aligned result $R_m$}

$\psi \leftarrow True$\label{alg1:ini_s}\\
$W \leftarrow -1$\label{alg1:ini_e}\\
\For{$i \leftarrow 1$ \KwTo $2^{|R_c|}-1$}{\label{alg1:traverse}
  $S \leftarrow \{\}$\label{alg1:compose_s}\\
  \For{$j \leftarrow 0$ \KwTo $K-1$}{
    \If{the $j$-th bit of $binary(i)=1$}{
      $S \leftarrow S \cup R_c[j]$\label{alg1:compose_e}
    }
  }
  \For{$i \leftarrow 1$ \KwTo $|S|$}{\label{alg1:check_s}
    \For{$j \leftarrow i$ \KwTo $|S|$}{
      \If{$S_i\asymp S_j$}{
        $\psi \leftarrow False$\label{alg1:check_e}
      }
    }
  }
  \If{$\psi=True$ \textbf{and} $\Delta(S,M) \leq \delta$}{\label{alg1:ans_s}
    \If{$\sum_{r \in S}W(r) > W$}{
    $R_m \leftarrow S$\\
    $W \leftarrow \sum_{r \in S}W(r)$\label{alg1:ans_e}
    }
  }
}
\Return{$R_m$}
\end{algorithm}

Time complexity of weight calculation is $O(|R_c|m^2)$ and Algorithm \ref{pseudo-exact} is $O(\sum_{i=1}^{2^{|R_c|}}s_i^2\times m)$, in which $s_i$ denotes the size of each composed time series. The worst-case complexity is $O(2^{|R_c|}|R_c|^2m)$.

\section{WEIGHTED SET PACKING-BASED APPROXIMATION}\label{approximation 1}

The composing algorithm in Section~\ref{exact} incurs high computational cost, as its time complexity involves enumerating all valid tuple combinations from the candidate set, which grows exponentially with the number of variables and the size of $R_c$. This is mainly because it will take composed time series with low sum of weight into account. So in the following section, we propose an alignment algorithm with lower time complexity following weighted k-set packing structure \cite{BestImp}. The candidate generation method is the same as Section \ref{candidate}. The optimization objectives of this approximate method is denoted as $W_M$.

In the process of time series composing, we also aim to choose several non-conflicting tuples in $R_c$ with maximal sum of weights. First, we initialize the tuple set $R_f$ by selecting tuples that are not conflicted with each other with greedy strategy. Next, for some $x \in R_f$, consider tuple set $Q$ which satisfies: (1)$\forall q_1,q_2\in Q, q_1 \centernot\asymp q_2$ and (2)$\forall q \in Q, q\asymp x$.

Choose all tuples that conflict with any tuple from $Q$ in $R_{f}$ to form a new tuple set $I$. We find the $Q$ which maximize $\frac{\sum_{q \in Q}W(q)}{\sum_{i \in I}W(i)}$, and then replace $I$ with $Q$. If multiple sets achieve the same maximal ratio, all such branches are preserved and explored.

This process stops when there is no such $Q$ can make $\sum_{x \in R_f}W(x)$ larger. Then, we use MARSS \cite{MARSS} model to calculate the normalized loss of all the composed time series, and choose the one with the largest sum of weights if model constraint is fulfilled as aligned result $R_m$. Time complexity of each branch is $O(|R_c|m^2 K^m)$, $K$ is the maximal sum of tuples that each tuple conflicted with.
\begin{proposition}
Given input time series data and fixed constraints $\theta$, $\beta$ and $\delta$, approximation ratio $\xi=\frac{W_M}{W_O}$ of weighted set packing-based approximation satisfies $\xi$ no less than $\frac{3}{2(m+1)}$.
\end{proposition}

\section{GREEDY AND EXPECTATION-BASED APPROXIMATION}\label{approximation 2}

Even though the time complexity is slightly improved in Section \ref{approximation 1}, it still incurs high computational cost which grows exponentially with $m$ due to conflict checks during tuple selection. To address this, we propose two more efficient approximation-based composing methods in this section, aiming to reduce computation while preserving alignment quality.

The two methods are inspired by different heuristic strategies: a greedy-based method that iteratively selects high-weight tuples, and an expectation-based method that estimates the potential contribution of each tuple afterwards. They differ in practical suitability for different data characteristics. Specifically, the Greedy-based method is highly efficient and performs well when candidate weights are relatively balanced and missing rates are low, as it quickly identifies high-weight tuples. In contrast, the Expectation-based strategy estimates the potential contribution of subsequent tuples, making it more robust in scenarios with high missing rates or skewed candidate weights, where local decisions are more likely to be suboptimal.

The candidate generation step remains the same as Section~\ref{candidate}, ensuring that all candidate tuples $r \in R_c$ satisfy the time and position constraints. The weight of each tuple is still computed as defined in Definition~\ref{def:weight}, serving as a key metric in both approximation strategies. The optimization objectives of the Greedy-based approximate method is denoted as $W_G$, and the optimization objectives of the Expectation-based approximate method is denoted as $W_E$.
\subsection{Greedy}

The Greedy-based strategy serves as a fast approximation. It iteratively selects the feasible tuple with local highest weight from the candidate set. Specifically, we iterate tuples in $R_c$ and continuously add each of them to a new tuple set $G$ if $G$ satisfies $\forall g_1,g_2 \in G, g_1\asymp g_2$ after this appending operation. This process stops when $G$ cannot fulfill this requirement if any tuple is added into it. Then, we pick out $g \in G$ with the largest weight and add it into the composed time series set $R_m$ as one tuple in aligned result. According to the experiment, it is worth mentioned that if there are several tuples with the same greatest weight, we can pick one of them randomly. After that, we continue adding tuples into $G$ in the same way.

This process stops when all tuples in $R_c$ is traversed. At last, the same as mentioned in the former part, we use MARSS \cite{MARSS} model to predict the values and check whether each of the composed $R_m$ satisfies model constraint $\delta$.

\begin{algorithm}[ht]
\caption{Time Series Compose-Greedy ($R_c$, $\delta$)}\label{pseudo-greedy}
\KwIn{Candidate set $R_c$, model constraint $\delta$}
\KwOut{Aligned result $R_m$}

$G \leftarrow \{\}$\\
\While{True}{
  $R_m \leftarrow \emptyset$\\
  \For{$i \leftarrow 0$ \KwTo $|R_c|$}{
    \If{$\forall r\in R_m,R_c[i]\centernot\asymp r$}{
      \If{$G=\emptyset$ or $\forall g\in G,g \asymp R_c[i]$}{
        $G \leftarrow G \cup \{R_c[i]\}$\\
      }
      \Else{
        $R_m \leftarrow R_m \cup \arg\max_{g\in G} W(g)$\\
        $G \leftarrow \{\}$\\
      }
    }
  }
  \If{$\Delta(R_m,M)\leq \delta$}{
    \Return{$R_m$}
  }
}
\end{algorithm}

Time complexity of the composing method for each round is $O(|R_c|m|G|+|R_c|log|G|)$. The worst-case complexity is $O(|R_c|m(4\beta+1)^m+|R_c|mlog(4\beta+1))$.

\begin{proposition}
Given input time series data and fixed constraints $\theta$, $\beta$ and $\delta$, approximation ratio $\xi=\frac{W_G}{W_O}$ of greedy-based approximation satisfies $\xi$ no less than $\frac{\left \lceil \frac{N}{m}  \right \rceil \times W_{min}}{N \times W_{max}}$.
$N$ is the number of aligned tuples under the same $R_c$ in the exact solution. $W_{max}$ is the theoretical maximal weight and $W_{min}$ is the theoretical minimal weight. 
\end{proposition}

\subsection{Expectation}

Differently, when we pick one tuple from each tuple set $G$ with the maximal weight, we use the sum of weights as a measure to evaluate each tuple of $g\in G$, instead of their original weights. Specifically, for each tuple $g\in G$, we define tuple set $E$ that satisfies $\forall r \in E,r>g\land r\centernot\asymp g$.

It is hard to find all tuples that satisfies this requirement in $R_c$, while we only need to consider about tuples in $R_c$ that conflict with at least one tuple in $G$, for the other tuples make no difference to the tuples in $G$.
\begin{lemma}\label{lemma 1}
For each $G$, there exists tuple $r$ in $R_c$ that conflict with at least one tuple in $G$ only when $\forall i \in [1,m],r[i]\leq max+\beta$, where $\max = \max_{i \in [1,m],\ r' \in G} r'[i]$. 
\end{lemma}

According to Lemma \ref{lemma 1}, the time complexity can be largely reduced. Finally, we add the total weight of set $E$ onto each weight $W(g)$ in $G$, find one tuple with the greatest weight and add it into the composed time series set $R_m$. Lastly, we check whether each of the composed $R_m$ satisfies model constraint $\delta$. Time complexity of this time series composing method for each round is $O(|R_c|m(|G|+z)+|R_c|log|G|)$. $z$ denotes the size of set that conflict with at least one tuple in $G$. The worst-case complexity is $O(2|R_c|m(4\beta+1)^{m}+|R_c|mlog(4\beta+1))$.

\begin{proposition}
Given input time series data and fixed constraints $\theta$, $\beta$ and $\delta$, approximation ratio $\xi=\frac{W_E}{W_O}$ of expectation-based approximation satisfies $\xi$ no less than $min(\frac{(\left \lceil \frac{N}{m}  \right \rceil+1) \times W_{min}}{N\times W_{max}},\frac{\left \lceil \frac{N}{m}  \right \rceil\times W_{min}}{(N-1) \times W_{max}+W_{min}})$. $N$ is the number of aligned tuples under the same $R_c$ in the exact solution.  
\end{proposition}

\section{EXPERIMENTS}
In this section, we compare our method in Section \ref{approximation 2} with existing methods: DTW \cite{DTW}, DDTW \cite{DDTW} and CTW \cite{CTW} extended in Procrustes analysis framework \cite{Extend} for multiple variables aligning, and GTW \cite{Extend,GTW}, TTW \cite{TTW} and SAMC \cite{SAMC} over real-world datasets Telemetry, Household, Water and Air\footnote{https://github.com/fangfcg/SAMC} with missing rates from 10\% \textasciitilde 40\% generated by MCAR. There are 5000 tuples in Telemetry and Water including 4 time series, 6839 tuples in Household including 4 time series and 1000 tuples in Air including 11 time series. We evaluate each method by the sum of aligned tuples and alignment accuracy.

We adopt F1-score to evaluate the alignment accuracy by comparing the recovered variable pairs against the ground truth. $R_{\text{truth}}$ denote the set of true \((T, V)\) pairs in the original complete data, and let $R$ denote the set of \((T, V)\) pairs finally aligned in $R_m$, $\text{Recall} = \frac{|R_{\text{truth}} \cap R|}{|R_{\text{truth}}|}$ and $\text{Precision} = \frac{|R_{\text{truth}} \cap R|}{|R|}$.
\subsection{Comparison with Existing Methods}
\renewcommand{\arraystretch}{1.8}
\setlength{\arrayrulewidth}{0.6pt}
\begin{table}
    \centering
    \setlength{\tabcolsep}{1pt} 
    \captionsetup{skip=8pt}
    {\fontsize{4pt}{5pt}\selectfont
    \scalebox{0.85}[1.0]{%
    \begin{tabular}{c|c|c|c|c|c|c|c|c|c|c|c|c|c|c|c}
        \hline
        \multirow{2}{*}{Dataset} & \multirow{2}{*}{Miss.} 
        & \multicolumn{2}{c|}{DTW} 
        & \multicolumn{2}{c|}{DDTW} 
        & \multicolumn{2}{c|}{CTW} 
        & \multicolumn{2}{c|}{GTW} 
        & \multicolumn{2}{c|}{TTW} 
        & \multicolumn{2}{c|}{SAMC} 
        & \multirow{2}{*}{Greedy} 
        & \multirow{2}{*}{Expect} \\
        \cline{3-14}
        & & Timer & TimesNet & Timer & TimesNet & Timer & TimesNet & Timer & TimesNet & Timer & TimesNet & Timer & TimesNet & & \\
        \hline
        \multirow{4}{*}{Telemetry} 
            & 10\% & 0.527 & 0.527 & 0.215 & 0.213 & 0.261 & 0.254 & 0.197 & 0.201 & 0.188 & 0.167 & 0.837 & 0.830 & \textbf{0.990} & \textbf{0.996} \\
            & 20\% & 0.509 & 0.514 & 0.197 & 0.192 & 0.256 & 0.247 & 0.316 & 0.185 & 0.263 & 0.229 & 0.853 & 0.853 & \textbf{0.993} & \textbf{0.993} \\
            & 30\% & 0.505 & 0.499 & 0.181 & 0.178 & 0.263 & 0.280 & 0.232 & 0.346 & 0.499 & 0.499 & 0.841 & 0.842 & \textbf{0.982} & \textbf{0.998} \\
            & 40\% & 0.487 & 0.488 & 0.174 & 0.161 & 0.271 & 0.284 & 0.195 & 0.171 & 0.527 & 0.309 & 0.883 & 0.873 & \textbf{0.964} & \textbf{0.997} \\
        \cline{1-16}
        \multirow{4}{*}{Household} 
            & 10\% & 0.158 & 0.159 & 0.143 & 0.143 & 0.253 & 0.249 & 0.302 & 0.247 & 0.393 & 0.004 & 0.956 & 0.956 & \textbf{0.998} & \textbf{0.998} \\
            & 20\% & 0.149 & 0.148 & 0.126 & 0.127 & 0.236 & 0.246 & 0.689 & 0.856 & 0.002 & 0.003 & 0.960 & 0.961 & \textbf{0.994} & \textbf{0.998} \\
            & 30\% & 0.137 & 0.197 & 0.116 & 0.124 & 0.167 & 0.174 & 0.406 & 0.529 & 0.085 & 0.590 & 0.964 & 0.964 & \textbf{0.986} & \textbf{0.998} \\
            & 40\% & 0.128 & 0.143 & 0.117 & 0.119 & 0.058 & 0.062 & 0.300 & 0.195 & 0.187 & 0.113 & 0.963 & 0.964 & \textbf{0.972} & \textbf{0.998} \\
        \cline{1-16}
        \multirow{4}{*}{Water} 
            & 10\% & 0.168 & 0.168 & 0.106 & 0.109 & 0.079 & 0.076 & 0.105 & 0.120 & 0.007 & 0.001 & 0.941 & 0.942 & \textbf{0.992} & \textbf{0.995} \\
            & 20\% & 0.188 & 0.158 & 0.101 & 0.103 & 0.040 & 0.064 & 0.129 & 0.114 & 0.014 & 0.008 & 0.907 & 0.914 & \textbf{0.994} & \textbf{0.997} \\
            & 30\% & 0.005 & 0.003 & 0.098 & 0.099 & 0.055 & 0.058 & 0.213 & 0.138 & 0.016 & 0.047 & 0.913 & 0.926 & \textbf{0.985} & \textbf{0.997} \\
            & 40\% & 0.003 & 0.014 & 0.096 & 0.102 & 0.057 & 0.064 & 0.140 & 0.149 & 0.001 & 0.002 & 0.904 & 0.907 & \textbf{0.969} & \textbf{0.998} \\
        \cline{1-16}
        \multirow{4}{*}{Air} 
            & 10\% & 0.170 & 0.229 & 0.110 & 0.118 & 0.064 & 0.081 & 0.290 & 0.201 & 0.098 & 0.048 & 0.924 & 0.924 & \textbf{0.979} & \textbf{0.980} \\
            & 20\% & 0.222 & 0.200 & 0.104 & 0.113 & 0.063 & 0.070 & 0.286 & 0.386 & 0.123 & 0.184 & 0.907 & 0.907 & \textbf{0.982} & \textbf{0.984}\\
            & 30\% & 0.164 & 0.155 & 0.098 & 0.111 & 0.052 & 0.059 & 0.134 & 0.256 & 0.065 & 0.052 & 0.914 & 0.914 & \textbf{0.988} & \textbf{0.989}\\
            & 40\% & 0.193 & 0.161 & 0.096 & 0.117 & 0.054 & 0.067 & 0.439 & 0.363 & 0.032 & 0.049 & 0.907 & 0.907 & \textbf{0.986} & \textbf{0.989}\\
        \hline
    \end{tabular}
    }
    }
    \caption{F1-score performance comparison with the existing six methods on four missing datasets using time-series imputation models Timer and TimesNet. Each dataset contains missing rates from 10\% to 40\%.}
    \label{full_comparison}
\end{table}
\renewcommand{\arraystretch}{1.0}
\setlength{\arrayrulewidth}{0.6pt}

\renewcommand{\arraystretch}{1.2}
\begin{table}[htbp]
    \centering
    \setlength{\tabcolsep}{2.5pt}
    {\scriptsize
    \begin{tabular}{c|c|c|c|c|c}
        \hline
        \multirow{2}{*}{Dataset} & \multirow{2}{*}{Miss.} & \multicolumn{2}{c|}{SAMC} & \multirow{2}{*}{Greedy} & \multirow{2}{*}{Expectation} \\
        \cline{3-4}
                &       & Timer & TimesNet &        &         \\
        \hline
        \multirow{4}{*}{Telemetry} 
            & 10\% & 4885 & 4879 & 4946 & \textbf{4968} \\
            & 20\% & 4899 & 4902 & 4946 & \textbf{4978} \\
            & 30\% & 4927 & 4932 & 4905 & \textbf{4985} \\
            & 40\% & 4953 & 4954 & 4847 & \textbf{4990} \\
        \hline
        \multirow{4}{*}{Household}
            & 10\% & 6742 & 6742 & 6811 & \textbf{6820} \\
            & 20\% & 6759 & 6762 & 6788 & \textbf{6829} \\
            & 30\% & 6769 & 6768 & 6744 & \textbf{6831} \\
            & 40\% & 6778 & 6781 & 6672 & \textbf{6833} \\
        \hline
        \multirow{4}{*}{Water}
            & 10\% & 4874 & 4879 & 4964 & \textbf{4986} \\
            & 20\% & 4791 & 4810 & 4955 & \textbf{4992} \\
            & 30\% & 4830 & 4859 & 4924 & \textbf{4996} \\
            & 40\% & 4816 & 4805 & 4866 & \textbf{4996} \\
        \hline
        \multirow{4}{*}{Air}
            & 10\% & 942 & 942 & 962 & \textbf{965} \\
            & 20\% & 936 & 936 & 966 & \textbf{973} \\
            & 30\% & 938 & 938 & 976 & \textbf{978} \\
            & 40\% & 936 & 936 & 978 & \textbf{986} \\
        \hline
    \end{tabular}
    }
    \vspace{1em}
    \caption{Aligned tuples comparison with SAMC, Greedy, and Expectation methods on four missing datasets using time-series imputation models Timer and TimesNet. Each dataset contains missing rates from 10\% to 40\%.}
    \label{tuples}
\end{table}

We apply Timer \cite{Timer} and TimesNet \cite{TimesNet} for imputing missing values, and use linear interpolation to fill in missing timestamps, as other baseline methods cannot directly handle incomplete inputs.

Due to the high time complexities of methods in Section \ref{exact} and \ref{approximation 1}, as shown in Table~\ref{full_comparison}, we only run the experiments on methods mentioned in Section \ref{approximation 2}. Our proposed methods consistently achieve better alignment accuracy across four datasets with missing rates ranging from 10\% to 40\%. Specifically, the performance of the Greedy-based strategy degrades as the missing rate increases. This is mainly because it relies heavily on the individual weights of candidate tuples and can be easily misled when information is sparse. In contrast, the Expectation-based strategy demonstrates more stable performance under increasing missing rates, although it may incur higher computational cost.

Table \ref{tuples} compares the number of aligned tuples by SAMC \cite{SAMC} and our two methods. Here we compare our methods only with SAMC because the DTW-based methods will always align tuples the same size as the input data. SAMC will possibly align more tuples when the missing rate grows, which is because the imputation of timestamps may make more tuples satisfies its own time constraint. We can see from this table that Greedy-based strategy aligns more tuples when the missing rate is low while its performance drops sharply when the rate increase, and Expectation-based strategy can always align more tuples regardless of missing rate. This is mainly because the Expectation-based method takes the potential contribution afterward, and out definition on weights of each tuple helps to boost positional consistency.

Figure \ref{F1-performance} in Appendix \ref{appendix} shows the alignment accuracy over different sizes of dataset Household with 20\% missing, varies from 1k to 6k. The other methods may be influenced largely by the dataset size, while our methods have better performance over all sizes of data.
\subsection{Parameter Determination}
In this section, we first analyze how the time constraint $\theta$, position constraint $\beta$, model constraint $\delta$ and the other weight factors affect the alignment performance on both Greedy and Expectation methods. Then, we give introduce automatic determination of constraints and weight factors without the ground truth.
\subsubsection{Varying $\theta$ and $\beta$.}

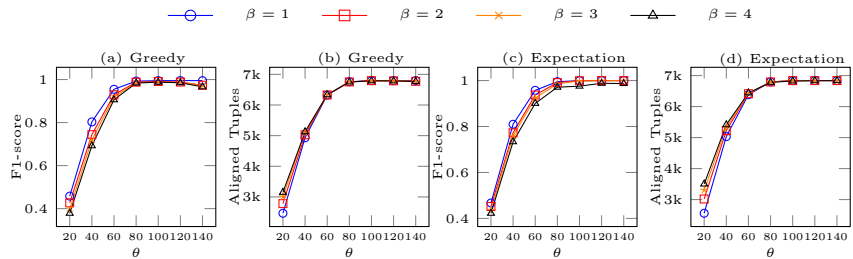
\begin{figure}[htbp]
  \centering
  \begin{tikzpicture}
    \begin{axis}[
        hide axis,
        xmin=0,
        xmax=1,
        ymin=0,
        ymax=1,
        legend columns=4,
        legend style={
          draw=none,
          font=\tiny,
          column sep=10pt,
          /tikz/every even column/.append style={column sep=10pt},
          at={(0.5,1.1)},
          anchor=south,
        }
      ]
      \addlegendimage{blue, mark=o}
      \addlegendentry{$\beta=1$}
      \addlegendimage{red, mark=square}
      \addlegendentry{$\beta=2$}
      \addlegendimage{orange, mark=x}
      \addlegendentry{$\beta=3$}
      \addlegendimage{black, mark=triangle}
      \addlegendentry{$\beta=4$}
    \end{axis}
  \end{tikzpicture}

  \begin{subfigure}[b]{3.7cm}
    \captionsetup{labelfont={tiny}, textfont={tiny}, justification=centering,skip=2pt}
    \caption{Greedy}\vspace{-1pt}
    \begin{tikzpicture}
      \begin{axis}[
          width=\linewidth,
          height=3.7cm,
          xlabel={\tiny $\theta$},
          xlabel style={font=\tiny, yshift=3pt},
          ylabel={\tiny F1-score},
          ylabel style={font=\tiny, yshift=-5pt},
          xtick={1,2,3,4,5,6,7},
          xticklabels={20, 40, 60, 80, 100, 120, 140},
          xticklabel style={font=\tiny, scale=0.8},
          ytick={0.2,0.4,0.6,0.8,1},
          yticklabels={0.2, 0.4, 0.6, 0.8, 1},
          tick label style={font=\tiny},
          label style={font=\tiny},
        ]
        \addplot[blue, mark=o, mark size=1.5pt] coordinates {
          (1,0.45806037967273433) (2,0.8032223983095615)
          (3,0.95475683770789) (4,0.9927573620745667)
          (5,0.9948729677702174) (6,0.9952364620250752)
          (7,0.9951029896533984)};
        \addplot[red, mark=square, mark size=1.5pt] coordinates {
          (1,0.4267726617937363) (2,0.7439236111111112)
          (3,0.9294910856446446) (4,0.9879591758724819)
          (5,0.9911197499809437) (6,0.9883774102583646)
          (7,0.9736435881310389)};
        \addplot[orange, mark=x, mark size=1.5pt] coordinates {
          (1,0.3985742903972735) (2,0.7185226636821489)
          (3,0.9185516808042726) (4,0.9847870957877838)
          (5,0.9882003088125965) (6,0.9877855904266467)
          (7,0.972323879231473)};
        \addplot[black, mark=triangle, mark size=1.5pt] coordinates {
          (1,0.3783714687460053) (2,0.6921347352191696)
          (3,0.905967485733336) (4,0.9838333651824958)
          (5,0.9880505422042652) (6,0.9851155876579444)
          (7,0.9661995996568487)};
      \end{axis}
    \end{tikzpicture}
  \end{subfigure}
  \hspace{-1cm}
  \begin{subfigure}[b]{3.7cm}
    \captionsetup{labelfont={tiny}, textfont={tiny}, justification=centering}
    \caption{Greedy}\vspace{-3pt}
    \begin{tikzpicture}
      \begin{axis}[
          width=\linewidth,
          height=3.7cm,
          xlabel={\tiny $\theta$},
          xlabel style={font=\tiny, yshift=3pt},
          ylabel={\tiny Aligned Tuples},
          ylabel style={font=\tiny, yshift=-6pt},
          xtick={1,2,3,4,5,6,7},
          xticklabels={20, 40, 60, 80, 100, 120, 140},
          xticklabel style={font=\tiny, scale=0.8},
          ytick={1,2,3,4,5,6,7},
          yticklabels={1k, 2k, 3k, 4k, 5k, 6k, 7k},
          tick label style={font=\tiny},
          label style={font=\tiny},
        ]
        \addplot[blue, mark=o, mark size=1.5pt] coordinates {
          (1,2.467) (2,4.931) (3,6.324) (4,6.759) (5,6.795) (6,6.797) (7,6.796)};
        \addplot[red, mark=square, mark size=1.5pt] coordinates {
          (1,2.782) (2,5.028) (3,6.328) (4,6.753) (5,6.794) (6,6.789) (7,6.773)};
        \addplot[orange, mark=star, mark size=1.5pt] coordinates {
          (1,3.006) (2,5.098) (3,6.332) (4,6.749) (5,6.786) (6,6.79) (7,6.774)};
        \addplot[black, mark=triangle, mark size=1.5pt] coordinates {
          (1,3.15) (2,5.128) (3,6.332) (4,6.751) (5,6.79) (6,6.787) (7,6.768)};
      \end{axis}
    \end{tikzpicture}
  \end{subfigure}
  \hspace{-1.2cm}
  \begin{subfigure}[b]{3.7cm}
    \captionsetup{labelfont={tiny}, textfont={tiny}, justification=centering}
    \caption{Expectation}\vspace{-1pt}
    \begin{tikzpicture}
      \begin{axis}[
          width=\linewidth,
          height=3.7cm,
          xlabel={\tiny $\theta$},
          xlabel style={font=\tiny, yshift=3pt},
          ylabel={\tiny F1-score},
          ylabel style={font=\tiny, yshift=-5pt},
          xtick={1,2,3,4,5,6,7},
          xticklabels={20, 40, 60, 80, 100, 120, 140},
          xticklabel style={font=\tiny, scale=0.8},
          ytick={0.2,0.4,0.6,0.8,1},
          yticklabels={0.2, 0.4, 0.6, 0.8, 1},
          tick label style={font=\tiny},
          label style={font=\tiny},
        ]
        \addplot[blue, mark=o, mark size=1.5pt] coordinates {
          (1,0.467020805666224) (2,0.8097823707141918)
          (3,0.9573593882084651) (4,0.995265006873377)
          (5,0.9994858611825193) (6,1) (7,1)};
        \addplot[red, mark=square, mark size=1.5pt] coordinates {
          (1,0.4520014728315186) (2,0.7722469709686444)
          (3,0.9370906381142878) (4,0.9909885829928596)
          (5,0.9992383853769993) (6,1) (7,1)};
        \addplot[orange, mark=star, mark size=1.5pt] coordinates {
          (1,0.4357433313725992) (2,0.7599693577903562)
          (3,0.9237719023779724)(4,0.98581600901055699)
          (5,0.9982864976010967) (6,0.9991051880057116) (7,0.9995621466237697)};
        \addplot[black, mark=triangle, mark size=1.5pt] coordinates {
          (1,0.4223537080679937) (2,0.7336479251751064)
          (3,0.9006345797129747) (4,0.9710106179818196)
          (5,0.9757463752929296) (6,0.9881945237823222) (7,0.9870189196391184)};
      \end{axis}
    \end{tikzpicture}
  \end{subfigure}
  \hspace{-1cm}
  \begin{subfigure}[b]{3.7cm}
    \captionsetup{labelfont={tiny}, textfont={tiny}, justification=centering}
    \caption{Expectation}\vspace{-3pt}
    \begin{tikzpicture}
      \begin{axis}[
          width=\linewidth,
          height=3.7cm,
          xlabel={\tiny $\theta$},
          xlabel style={font=\tiny, yshift=3pt},
          ylabel={\tiny Aligned Tuples},
          ylabel style={font=\tiny, yshift=-6pt},
          xtick={1,2,3,4,5,6,7},
          xticklabels={20, 40, 60, 80, 100, 120, 140},
          xticklabel style={font=\tiny, scale=0.8},
          ytick={1,2,3,4,5,6,7},
          yticklabels={1k, 2k, 3k, 4k, 5k, 6k, 7k},
          tick label style={font=\tiny},
          label style={font=\tiny},
        ]
        \addplot[blue, mark=o, mark size=1.5pt] coordinates {
          (1,2.559) (2,5.032) (3,6.384) (4,6.795) (5,6.837) (6,6.839) (7,6.839)};
        \addplot[red, mark=square, mark size=1.5pt] coordinates {
          (1,3.02) (2,5.222) (3,6.414) (4,6.795) (5,6.837) (6,6.839) (7,6.839)};
        \addplot[orange, mark=star, mark size=1.5pt] coordinates {
          (1,3.307) (2,5.331) (3,6.427) (4,6.794) (5,6.837) (6,6.839) (7,6.839)};
        \addplot[black, mark=triangle, mark size=1.5pt] coordinates {
          (1,3.502) (2,5.415) (3,6.441) (4,6.782) (5,6.826) (6,6.835) (7,6.837)};
      \end{axis}
    \end{tikzpicture}
  \end{subfigure}

  \caption{Alignment performance for each $\theta$ with a determined $k_1=3$, $k_2=2$ and $b=1$, $c=1$ over dataset Household with 20\% missing.}
  \label{varying theta and beta}
\end{figure}

With fixed model constraint $\delta$ and set of weight factors $k_1,k_2,b,c$, we have position constraint $\beta$ varying from 1 to 4 and time constraint $\theta$ varying from 20 to 140 on dataset Household with 20\% missing. Figure \ref{varying theta and beta} reports the performance on each set of constraint. It is not surprising that with $\theta$ grows, both methods perform better because there are more candidate tuples that can satisfies time constraint, but the performance will be a bit worse when $\theta$ is too large. This is mainly because there are more misleading tuples in candidate. Also, the two methods perform stably when $\beta$ changes.
\subsubsection{Determining $\theta$ and $\beta$.}

To determine the time constraint $\theta$, following SAMC \cite{SAMC}, we collect non-missing timestamp differences between variable pairs that are recorded at the same semantic time but exhibit variations due to network or hardware delays. We then examine the distribution of these differences and set $\theta$ as a high-percentile threshold, such as the 95th percentile, in order to tolerate most of the observed delays.

For the position constraint $\beta$, which is unique to our formulation, we estimate it based on the structure of the candidate tuple set. We first specify an initial lower bound $\beta'$, and construct the candidate set $R_c$ using the method described in Section~\ref{candidate}. Then, we observe the distribution of line index distances between any two values in each tuple from candidate. Finally, we select the 80th percentile of this distribution as the position constraint $\beta$, under the condition that $\beta > \beta'$. This ensures that most compact and structurally consistent tuples are retained, while filtering out tuples with excessive positional gaps.
\subsubsection{Varying $\delta$ and $k_1,k_2$.}
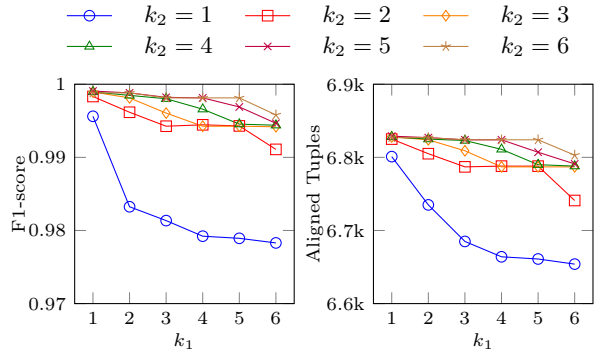
\begin{figure}[htbp]
  \centering
  \begin{tikzpicture}
    \begin{axis}[
        hide axis,
        xmin=0, xmax=1,
        ymin=0, ymax=1,
        legend columns=3,
        legend style={
          draw=none,
          font=\footnotesize,
          column sep=10pt,
          at={(0.5,1.15)},
          anchor=south
        }
      ]
      \addlegendimage{blue, mark=o}
      \addlegendentry{$k_2 = 1$}
      \addlegendimage{red, mark=square}
      \addlegendentry{$k_2 = 2$}
      \addlegendimage{orange, mark=diamond}
      \addlegendentry{$k_2 = 3$}
      \addlegendimage{green!50!black, mark=triangle}
      \addlegendentry{$k_2 = 4$}
      \addlegendimage{purple, mark=x}
      \addlegendentry{$k_2 = 5$}
      \addlegendimage{brown, mark=star}
      \addlegendentry{$k_2 = 6$}
    \end{axis}
  \end{tikzpicture}
  
  \begin{subfigure}[b]{4.5cm}
    \begin{tikzpicture}
      \begin{axis}[
          width=\linewidth,
          height=4.5cm,
          xlabel={\scriptsize $k_1$},
          xlabel style={font=\scriptsize, yshift=3pt},
          ylabel={\scriptsize F1-score},
          ylabel style={font=\scriptsize, yshift=-5pt},
          xtick={1,2,3,4,5,6},
          xticklabels={1,2,3,4,5,6},
          ymin=0.97,
          ymax=1,
          ytick={0.97,0.98,0.99,1},
          yticklabels={0.97,0.98,0.99,1},
          tick label style={font=\scriptsize},
          label style={font=\scriptsize},
        ]
        \addplot[blue, mark=o] coordinates {(1,0.9956143696133019) (2,0.9832442748091603) (3,0.9813503350067765) (4,0.9792179538558424) (5,0.9789272761977477) (6,0.9782977992403276)};
        \addplot[red, mark=square] coordinates {(1,0.9983232985290754) (2,0.9961684363026363) (3,0.9942593404916751) (4,0.9944506950933465) (5,0.9942981368828543) (6,0.9910913565174262)};
        \addplot[orange, mark=diamond] coordinates {(1,0.9989330081548663) (2,0.9981327642710159) (3,0.99605421170012) (4,0.9942593404916751) (5,0.9942593404916751) (6,0.9941830526576775)};
        \addplot[green!50!black, mark=triangle] coordinates {(1,0.9989520016767973) (2,0.9984755516597431) (3,0.9980182171576659) (4,0.9965882016582484) (5,0.9945455238967083) (6,0.9943931650011443)};
        \addplot[purple, mark=x] coordinates {(1,0.9990664354983139) (2,0.9988185525363009) (3,0.9982089779742397) (4,0.9981327642710159) (5,0.9968931668731535) (6,0.9946605644546148)};
        \addplot[brown, mark=star] coordinates {(1,0.9988185525363009) (2,0.9988185525363009) (3,0.9981325507831852) (4,0.9981327642710159) (5,0.9981327642710159) (6,0.9958056890109052)};
      \end{axis}
    \end{tikzpicture}
  \end{subfigure}
  \hspace{-0.8cm}
  \begin{subfigure}[b]{4.5cm}
    \begin{tikzpicture}
      \begin{axis}[
          width=\linewidth,
          height=4.5cm,
          xlabel={\scriptsize $k_1$},
          xlabel style={font=\scriptsize, yshift=3pt},
          ylabel={\scriptsize Aligned Tuples},
          ylabel style={font=\scriptsize, yshift=-5pt},
          xtick={1,2,3,4,5,6},
          xticklabels={1,2,3,4,5,6},
          ymin=6.6,
          ymax=6.9,
          ytick={6.6,6.7,6.8,6.9},
          yticklabels={6.6k,6.7k,6.8k,6.9k},
          tick label style={font=\scriptsize},
          label style={font=\scriptsize},
        ]
        \addplot[blue, mark=o] coordinates {(1,6.801) (2,6.735) (3,6.685) (4,6.664) (5,6.661) (6,6.654)};
        \addplot[red, mark=square] coordinates {(1,6.825) (2,6.805) (3,6.787) (4,6.788) (5,6.788) (6,6.741)};
        \addplot[orange, mark=diamond] coordinates {(1,6.828) (2,6.824) (3,6.809) (4,6.787) (5,6.787) (6,6.787)};
        \addplot[green!50!black, mark=triangle] coordinates {(1,6.828) (2,6.825) (3,6.823) (4,6.811) (5,6.79) (6,6.788)};
        \addplot[purple, mark=x] coordinates {(1,6.829) (2,6.827) (3,6.824) (4,6.824) (5,6.807) (6,6.791)};
        \addplot[brown, mark=star] coordinates {(1,6.827) (2,6.827) (3,6.824) (4,6.824) (5,6.824) (6,6.803)};
      \end{axis}
    \end{tikzpicture}
  \end{subfigure}
  \caption{Alignment performance of Greedy-based strategy for each $k$ with determined $\theta=90$, $\beta=1$ and $b=1$, $c=1$ over dataset Household with 20\% missing.}
  \label{varying k}
\end{figure}
$\delta$ is designed to filter the aligned results that are not consistent enough within it, so different $\delta$ will not largely influence the alignment performance, including F1-score and the number of aligned tuples. With fixed model constraint $\delta$, time constraint $\theta$, position constraint $\beta$ and $b,c$, Figure \ref{varying k} shows the performance of Greedy strategy over dataset Household with 20\% missing. The performance of Greedy strategy will be influenced by $k_1$ and $k_2$, but it can always maintain in a reasonable range which is F1-score no less than 0.97 and aligned tuples no less than 6.6k. Comparing to Greedy strategy, Expectation strategy perform much more robustly with F1-score no less than 0.99 and aligned tuples no less than 6.8k.
\subsubsection{Determining $\delta$ and $k_1,k_2,b,c$.}
Firstly, we can directly give $b=1$ and $c=1$ for these two weight factors are designed to make sure that no tuple gets a weight equals to zero. For each pair of $(k_1,k_2)$, we can get several aligned results regardless of model constraint. We evaluate each of them by $\Delta$ mentioned in Definition \ref{def:model}. Specifically, we observe the average of $\Delta$ of each aligned results under the same $(k_1,k_2)$ and select the one with minimal $\overline{\Delta}$. Then, we choose the corresponding $(k_1,k_2)$ as weight factors and have model constraint $\delta=\overline{\Delta}$.
\section{CONCLUSION}
In this paper, we present a constraint-based framework for aligning incomplete multivariate time series without imputation. The method applies time, position, and model-based constraints to ensure valid and consistent alignment. To address the computational challenge of exact alignment, we develop three approximation algorithms balancing efficiency and accuracy. Experiments on real-world datasets show our methods outperform existing baselines, particularly under high missing rates, while Greedy and Expectation strategies provide practical scalability with strong performance.
\clearpage
\bibliographystyle{splncs04}
\bibliography{reference}

\begin{thebibliography}{10}
\providecommand{\url}[1]{\texttt{#1}}
\providecommand{\urlprefix}{URL }
\providecommand{\doi}[1]{https://doi.org/#1}

\bibitem{intro_5}
Bastos, A.F., Santoso, S., Krishnan, V.K., Zhang, Y.: Machine learning based prediction of distribution network voltage and sensors allocation. Technical report, National Renewable Energy Laboratory (NREL) (2020)

\bibitem{BestImp}
Chandra, B., Halld{\'o}rsson, M.M.: Greedy local improvement and weighted set packing approximation. Journal of Algorithms  \textbf{39}(2),  223--240 (2001)

\bibitem{SAMC}
Fang, C., Song, S., Mei, Y., Yuan, Y., Wang, J.: On aligning tuples for regression. In: Proceedings of the 28th ACM SIGKDD Conference on Knowledge Discovery and Data Mining (KDD). pp. 336--346 (2022)

\bibitem{MARSS}
Holmes, E.E., Ward, E.J., Wills, K.: {MARSS}: Multivariate autoregressive state-space models for analyzing time-series data. The R Journal  \textbf{4}(1),  11--19 (2012)

\bibitem{DDTW}
Keogh, E.J., Pazzani, M.J.: Derivative dynamic time warping. In: Proceedings of the 2001 SIAM International Conference on Data Mining (SDM). pp. 1--11. Society for Industrial and Applied Mathematics (2001)

\bibitem{TTW}
Khorram, S., McInnis, M.G., Provost, E.M.: Trainable time warping: Aligning time-series in the continuous-time domain. In: Proceedings of the IEEE International Conference on Acoustics, Speech and Signal Processing (ICASSP). pp. 3502--3506 (2019)

\bibitem{DTW}
Li, S.Z., Jain, A.K.: Dynamic time warping ({DTW}). In: Encyclopedia of Biometrics, p.~231. Springer US (2009)

\bibitem{Timer}
Liu, Y., Zhang, H., Li, C., Huang, X., Wang, J., Long, M.: Timer: Generative pre-trained transformers are large time series models. In: Proceedings of the 41st International Conference on Machine Learning (ICML). vol.~235, pp. 32369--32399 (2024)

\bibitem{intro_2}
Marx, S.E., Luck, J.D., Pitla, S.K., Hoy, R.M.: Comparing various hardware/software solutions and conversion methods for controller area network ({CAN}) bus data collection. Computers and Electronics in Agriculture  \textbf{128},  141--148 (2016)

\bibitem{intro_1}
Philipp, P., Altmannshofer, S.: Experimental validation of a new moving horizon estimator approach for networked control systems with unsynchronized clocks. In: Proceedings of the American Control Conference (ACC). pp. 4939--4944. IEEE (2012)

\bibitem{intro_3}
Silva, I., Moody, G., Scott, D.J., Celi, L.A., Mark, R.G.: Predicting in-hospital mortality of {ICU} patients: The {PhysioNet/Computing in Cardiology Challenge 2012}. In: Computing in Cardiology. vol.~39, p.~245 (2012)

\bibitem{mobile}
Tong, Y., She, J., Ding, B., Wang, L., Chen, L.: Online mobile micro-task allocation in spatial crowdsourcing. In: 2016 IEEE 32nd International Conference on Data Engineering (ICDE). pp. 49--60 (2016). \doi{10.1109/ICDE.2016.7498228}

\bibitem{Imputation}
Wang, J., Du, W., Yang, Y., Qian, L., Cao, W., Zhang, K., Wang, W., Liang, Y., Wen, Q.: Deep learning for multivariate time series imputation: A survey (2025), \url{https://arxiv.org/abs/2402.04059}

\bibitem{TimesNet}
Wu, H., Hu, T., Liu, Y., Zhou, H., Wang, J., Long, M.: Timesnet: Temporal 2d-variation modeling for general time series analysis (2022), \url{https://arxiv.org/abs/2210.02186}

\bibitem{intro_4}
Ye, M., Yi, X., Jiao, S.: Energy optimization by parameter matching for a truck-mounted concrete pump. Energy Procedia  \textbf{88},  574--580 (2016)

\bibitem{CTW}
Zhou, F., la~Torre, F.D.: Canonical time warping for alignment of human behavior. In: Proceedings of the Neural Information Processing Systems (NeurIPS). pp. 2286--2294. Curran Associates, Inc. (2009)

\bibitem{Extend}
Zhou, F., la~Torre, F.D.: Generalized time warping for multi-modal alignment of human motion. In: Proceedings of the IEEE Conference on Computer Vision and Pattern Recognition (CVPR). pp. 1282--1289. IEEE Computer Society (2012)

\bibitem{GTW}
Zhou, F., la~Torre, F.D.: Generalized canonical time warping. IEEE Transactions on Pattern Analysis and Machine Intelligence  \textbf{38}(2),  279--294 (2016)

\end{thebibliography}
\clearpage
\appendix
\section{PROOFS}
\subsection{Proof of Proposition 1}
The time series composing procedure is equivalent to the weighted k-Set Packing (k-SP) problem with $m=k$. Given a ground set $V$ and a collection $S$ of sets, each of them contains $k$ elements from $V$, weighted k-SP problem is to find a subcollection of $S$ with maximized total weight so they are pairwise disjoint. In our scenario, let the set of all single tuples from the original data be $V$, and let $R_c$ correspond to $S$. The two problems are equivalent since they both maximize the sum of weights. Our time series composing process follows the framework of Weighted Set Packing Approximation \cite{BestImp}, which is proved to have a bounded approximation ratio.
\subsection{Proof of Proposition 2}
With greedy strategy, we have $\forall g_1,g_2 \in G, g_1\asymp g_2$ for each $G$, so each tuple $r$ in $R_c$ satisfies either $r\in R_m$ or $\exists r_m\in R_m, r_m \asymp r$. In another word, there must be at least one tuple in $G$ that conflict with it. Assuming that the exact solution contains $N$ tuples, denoted as $T=\left \{ T_1,T_2...T_{N} \right \}$. According to principle (1) in Problem Statement, we have $\forall t_1,t_2 \in T, t_1 \centernot\asymp t_2$, so each tuple in the approximation solution can be conflicted with $m$ tuples in $T$ at most, because each tuple can only conflict with one tuple in $T$ in one column. Therefore, we prove that $\left | R_m \right | \ge \left \lceil \frac{N}{m}  \right \rceil$.

We have
\begin{align}
\left\{
\begin{aligned}
&W_{max}=\frac{k_1\times p_{max}+b}{k_2\times d_{min}+c} \\
&W_{min}=\frac{k_1\times p_{min}+b}{k_2\times d_{max}+c}\\
\end{aligned}
\right.
\end{align}
$p_{max}$, $p_{min}$ are the maximal and minimal pairs in aligned tuples and $d_{max}$, $d_{min}$ are the maximal and minimal total distances in aligned tuples. 

To construct the minimum of approximation weight, we divide $T$ into several sets with $m$ tuples in each of them. For each of the set $S=\left \{ S_1 ,S_2...S_m\right \}$, we construct two tuples $m_1, m_2$ in candidate that satisfy: (1) $\forall s \in S,m_1,m_2<s$; (2) $m_1 \asymp m_2$; (3) $\forall s \in S,m_1\asymp s$ and $m_2 \centernot\asymp S_1$; (4) $W(m_1)=W(m_2)=W_{min}$ and $\forall s\in S,W(s)=W_{max}$. In which case, we can get the minimal sum of weights.

Therefore, the minimum approximation rate bound guarantee is $\frac{\left \lceil \frac{N}{m}  \right \rceil \times W_{min}}{{N} \times W_{max}}$, which is the theoretical minimum.

\subsection{Proof of Lemma 1}
For each tuple set $G$ with its maximum $max$ among all columns, denoted as $\forall g\in G, i\in[1,m], g[i]\leq max$, if any $r[i] > max+\beta$, the minimal maximum of $r$ among all columns is $max+\beta+1$, and its minimal minimum is $max+1$, according to Definition \ref{def:position}, $\forall i,j \in[1,m],|I[(T_i,V_i)]-I[(T_j,V_j)]|\leq \beta$. So $\forall i\in[1,m], r[i]\in[max+1, max+\beta+1]$. Therefore, there is always $r\centernot\asymp g$.
\subsection{Proof of Proposition 3}
Denote the aligned result of expectation approximation as $R_m=\left \{E_1,E_2...E_l\right \}$. The same with Greedy strategy, we also guarantee that the Expectation approximation can also align tuples no less than $\left \lceil \frac{N}{m}  \right \rceil$. We assume that weights $W(E_1)=W(E_2)=...=W(E_{l-1})=W_{min}$ can be got, which is the theoretical minimum. For the last tuple $E_l$, here are the following two cases:

1. $E_l$ is in the last $G$.

This condition exists only when $E_l$ is conflicted with one tuple in $S$. If $E_l$ is finally selected, there must be $W(E_l) \ge W(S)$. When $W(E_l)=W(S)=W_{min}$, the approximation gets its minimum. Therefore, we have bound guarantee $\frac{\left \lceil \frac{N}{m}  \right \rceil\times W_{min}}{(N-1) \times W_{max}+W_{min}}$.

2. $E_l$ is not in the last $G$.

There will be at least one tuple $f_1 \in R_c$ not conflict with at least one tuple $f_2 \in G$, according to \ref{approximation 2}. So if we finally choose $E_l$, there must be (1) $W(E_l)\ge 2W_{min}$ or (2) $\exists f\in R_c, f>E_l$ and $f\centernot\asymp E_l$. Therefore, in both cases, we can get the minimum approximation rate bound guarantee $\frac{(\left \lceil \frac{N}{m}\right \rceil +1)\times W_{min}}{N \times W_{max}}$.

\section{EXPERIMENT}\label{appendix}
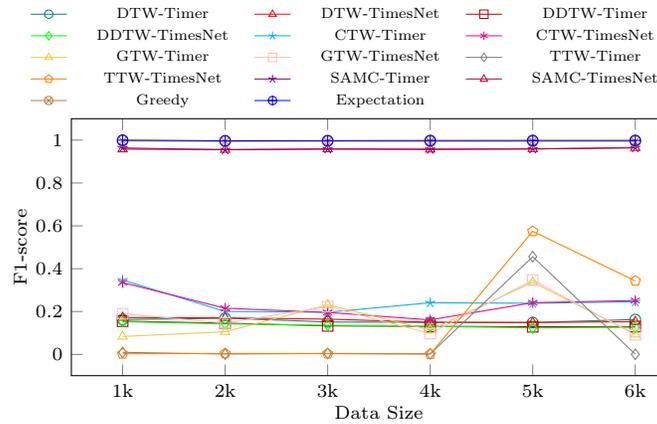
\begin{figure}[htbp]
  \centering
  \begin{tikzpicture}[scale=0.8]
    \begin{axis}[
        hide axis,
        xmin=0, xmax=1,
        ymin=0, ymax=1,
        legend columns=3,
        legend style={
          draw=none,
          font=\scriptsize,
          column sep=12pt,
          /tikz/every even column/.append style={column sep=10pt},
          at={(0.5,1.15)},
          anchor=south
        }
      ]
      \addlegendimage{teal, mark=o}
      \addlegendentry{DTW-Timer}
      \addlegendimage{red, mark=triangle}
      \addlegendentry{DTW-TimesNet}
      \addlegendimage{red!70!black, mark=square}
      \addlegendentry{DDTW-Timer}
      \addlegendimage{green, mark=diamond}
      \addlegendentry{DDTW-TimesNet}
      \addlegendimage{cyan, mark=star}
      \addlegendentry{CTW-Timer}
      \addlegendimage{magenta, mark=asterisk}
      \addlegendentry{CTW-TimesNet}
      \addlegendimage{yellow!60!brown, mark=triangle}
      \addlegendentry{GTW-Timer}
      \addlegendimage{pink, mark=square}
      \addlegendentry{GTW-TimesNet}
      \addlegendimage{gray, mark=diamond}
      \addlegendentry{TTW-Timer}
      \addlegendimage{orange, mark=pentagon}
      \addlegendentry{TTW-TimesNet}
      \addlegendimage{violet, mark=star}
      \addlegendentry{SAMC-Timer}
      \addlegendimage{purple, mark=triangle}
      \addlegendentry{SAMC-TimesNet}
      \addlegendimage{brown, mark=otimes}
      \addlegendentry{Greedy}
      \addlegendimage{blue, mark=oplus}
      \addlegendentry{Expectation}
    \end{axis}
  \end{tikzpicture}
    \begin{tikzpicture}
      \begin{axis}[
          width=0.8\linewidth,
          height=5cm,
          xlabel={\scriptsize Data Size},
          xlabel style={font=\scriptsize, yshift=3pt},
          ylabel={\scriptsize F1-score},
          ylabel style={font=\scriptsize, yshift=-3pt},
          xtick={1,2,3,4,5,6},
          xticklabels={1k, 2k, 3k, 4k, 5k, 6k},
          ytick={0,0.2,0.4,0.6,0.8,1},
          yticklabels={0, 0.2, 0.4, 0.6, 0.8, 1},
          tick label style={font=\scriptsize},
          label style={font=\scriptsize},
        ]
        \addplot[teal, mark=o] coordinates{
            (1,0.16250739207569487)(2,0.16920415224913493)(3,0.15148968678380442)(4,0.1494262790095764)(5,0.1487359291382174)(6,0.16272662466524093)};
        \addplot[red, mark=triangle] coordinates{
            (1,0.17188315075639019)(2,0.17032586087358448)(3,0.16452867665314905)(4,0.15020013246177327)(5,0.14904167723781628)(6,0.15267557884193086)};
        \addplot[red!70!black, mark=square] coordinates{
            (1,0.15497270669684463)(2,0.14590609235545204)(3,0.13295663713025985)(4,0.13082730923694783)(5,0.12910543954627482)(6,0.1299933170931511)};
        \addplot[green, mark=diamond] coordinates{
            (1,0.15261780104712044)(2,0.14272809394760613)(3,0.13576598569531886)(4,0.13087474702046323)(5,0.12389654638376955)(6,0.1265005808700142)};
        \addplot[cyan, mark=star] coordinates{
            (1,0.34700043159257665)(2,0.20051077205160106)(3,0.19663328361389304)(4,0.2410937800885808)(5,0.23869782630880704)(6,0.245713320381789)};
        \addplot[magenta, mark=asterisk] coordinates{
            (1,0.3351480547072102)(2,0.21547360809833696)(3,0.19583264154076038)(4,0.16112909816741586)(5,0.24137495898846628)(6,0.2514030585633515)};
        \addplot[yellow!60!brown, mark=triangle] coordinates{
            (1,0.08355722729715206)(2,0.10580560991519895)(3,0.2313288278593259)(4,0.11414504583292845)(5,0.337038290340214)(6,0.093971092495334)};
        \addplot[pink, mark=square] coordinates{
            (1,0.1895975397780452)(2,0.14634467329078107)(3,0.22314014026935156)(4,0.09725062485798683)(5,0.3470621969657473)(6,0.09178879916870522)};
        \addplot[gray, mark=diamond] coordinates{
            (1,0.009557618787547788)(2,0.0016794780698921258)(3,0.004707694941367799)(4,0.002535336248970019)(5,0.4561056277507226)(6,0.0006991260923845194)};
        \addplot[orange, mark=pentagon] coordinates{
            (1,0.002752064048036027)(2,0.004431122116512446)(3,0.0032521716804313405)(4,0.0012503516614047702)(5,0.5745911032399208)(6,0.3431285376591269)};
        \addplot[violet, mark=star] coordinates{
            (1,0.963483905869624)(2,0.9549656629688326)(3,0.9586827127775123)(4,0.95745794881864)(5,0.9588523773307807)(6,0.9642381793982239)};
        \addplot[purple, mark=triangle] coordinates{
            (1,0.9577693557119654)(2,0.9559648775335049)(3,0.9583604706699664)(4,0.9571246361644373)(5,0.9590857861767096)(6,0.9644610249361869)};
        \addplot[brown, mark=otimes] coordinates{
            (1,0.9961047683008731)(2,0.9942249639060244)(3,0.9947684724804359)(4,0.9942049746060685)(5,0.9944431400172185)(6,0.9945447827693378)};
        \addplot[blue, mark=oplus] coordinates{
            (1,1.0)(2,0.9969182348698445)(3,0.9975807845170209)(4,0.9981781508230854)(5,0.998462139971328)(6,0.9985232800555942)};
      \end{axis}
    \end{tikzpicture}
  \caption{Comparing with the existing alignment methods over dataset Household with 20\% missing on F1-score of alignment results.}
  \label{F1-performance}
\end{figure}

\end{document}